\documentclass[12pt]{article}
\usepackage{graphics}
\begin{document}
\setlength{\baselineskip}{3.0ex}
 
\rightline{SLAC-PUB-7682}
\rightline{December 1, 1997}
\begin{center}
\vspace*{1.8cm}
{\large\bf Search for \mbox{\boldmath $CP$} violation and 
\mbox{\boldmath $b\rightarrow sg$}\\
in inclusive \mbox{\boldmath $B$}
\renewcommand{\thefootnote}{\fnsymbol{footnote}}decays\footnote{Work 
supported by U.S. Department of Energy contract DE-AC03-76SF00515.}}\\
\vspace*{6.0ex}
{\large Mourad Daoudi}\\
\vspace*{1.5ex}
{\it Stanford Linear Accelerator Center\\
Stanford, CA 94309}\\
\vspace*{0.4in}
Representing the SLD Collaboration\\
\vspace*{1.2in}
{\bf Abstract}\\
\end{center}
\noindent
We present preliminary results on two analyses performed by the SLD 
Collaboration using inclusive $B$ decays: a search for CP violation and a 
search for the $b\rightarrow sg$ transition. 

\vspace*{2.2in}
\begin{center}
{\footnotesize
Talk given at the International Europhysics Conference on High Energy Physics\\
Jerusalem, Israel, August 19--26, 1997}
\end{center}
 
\newpage
\pagestyle{plain}

%
% ------ CP violation
%
\section{Search for \mbox{\boldmath $CP$} violation in inclusive 
\mbox{\boldmath $B$} decays}
Because they involve large branching fractions and sizable CP asymmetries,
(semi-) inclusive $B$ decays have been proposed\cite{dunietz} as a means of
searching for CP violation, and extracting measurements of CKM parameters.
The totally inclusive asymmetry provides a measurement of the CP observable
$a = {\cal I}m{\Gamma_{12}\over M_{12}}$. It is the focus of this analysis.
Its time dependence is\cite{dunietz}:
{\small
$$
{\cal{A}}(t) = 
{{\Gamma(B^0(t)\rightarrow all)-\Gamma(\bar{B^0}(t)\rightarrow all)}\over 
 {\Gamma(B^0(t)\rightarrow all)+\Gamma(\bar{B^0}(t)\rightarrow all)}}
= {a \left({{\Delta m\ \tau_B}\over 2} \sin\Delta m\ t 
- \sin^2{{\Delta m\ t}\over 2}\right)}\cdot
$$ }
A non-zero value of $a$ implies CP violation. Due to the large value of 
$\Delta m_s$, this analysis is only sensitive to asymmetries in $B_d$ decays
for which $a_d$ is expected to be $\approx 10^{-3}$ in the Standard Model.\\
\begin{center}
\includegraphics{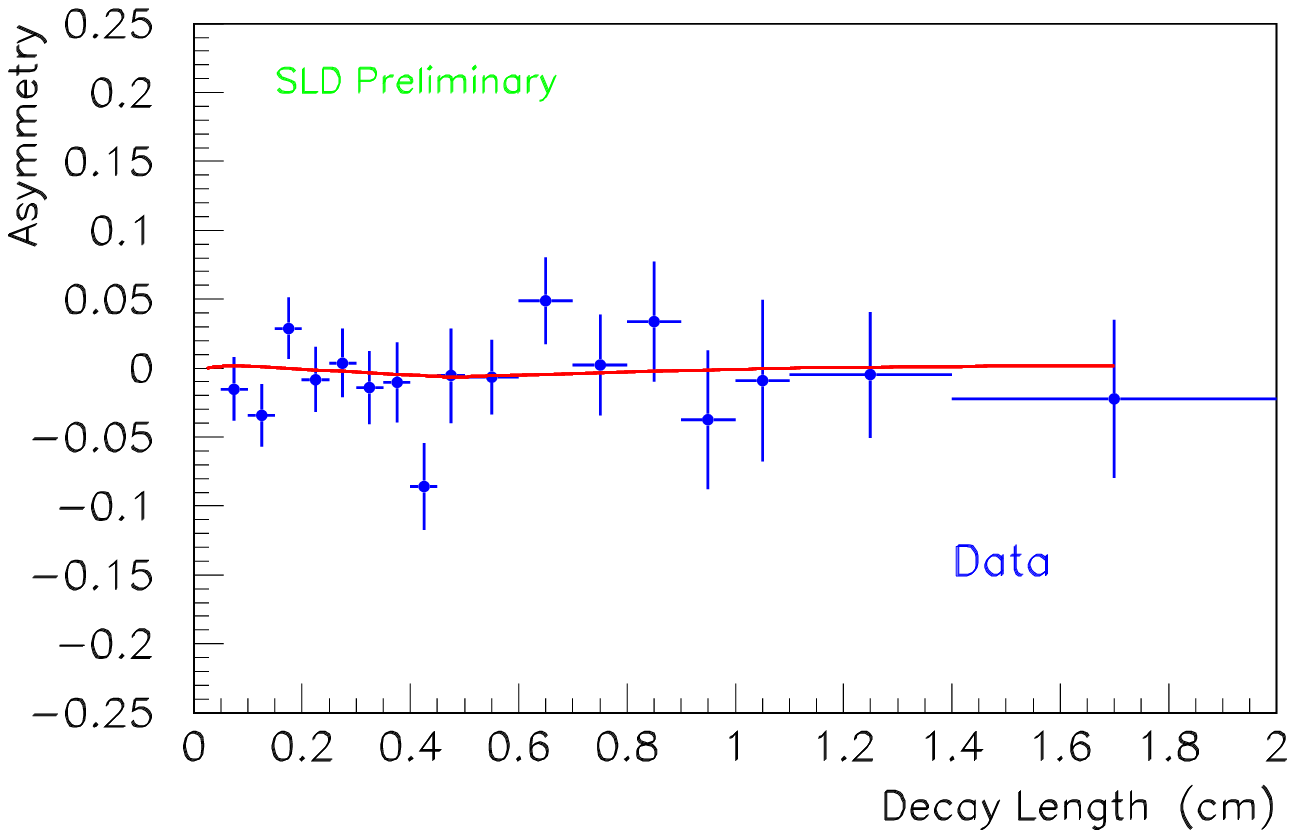}\\
{\small{\bf Fig.1.} Asymmetry as a function of decay length.}
\end{center}
\par $B$-decay vertices are reconstructed using a topological
technique\cite{jackson}. From the 1993-96 data sample  
($200k\ Z^0$'s), about $11k$ neutral and $19k$ charged vertices are
selected, with a $B_d$ content of 50\% and 35\%, respectively.
The crucial part of the analysis is 
tagging the flavor of the $b$ quark at production. This is done mainly 
using the left-right forward-backward asymmetry (given by the
electron beam longitudinal polarization and the thrust axis polar angle) 
and the opposite hemisphere momentum-weighted jet charge.
These two tags which have an efficiency of 100\%,
are complemented by additional information from the opposite hemisphere when
it is available (vertex charge, sign of a high-$p_T$ lepton, charge sum for
kaons from a $B$-decay). The overall $b$-flavor tag 
purity is estimated to be 84\%.  
\par The measured asymmetry is shown in Fig.~1 as a
function of decay length. A binned $\chi^2$ fit is performed and a value
of $a_d = -0.04\pm 0.12(stat)\pm 0.05(sys)$ is obtained. This gives a 95\% C.L.
limit of $-0.29 < a_d < 0.22$. 
%
% b --> sg
%
\section{Search for enhanced \mbox{\boldmath $b\rightarrow sg$} in 
inclusive \mbox{\boldmath $B$} decays}
It was suggested recently that a branching ratio $\sim 10\%$ for the
$b\rightarrow sg$ transition (0.2\% in the Standard Model) could resolve a 
variety of $B$ decay puzzles
(e.g., $b$ semileptonic branching ratio, number of $c$ quarks produced per
$B$ decay, etc)\cite{kagan}. The search strategy consists of looking for
an excess in kaon production at high $p_T$, where the signal-to-background 
ratio is expected to be of the order of 1:1 (for a 10\% branching ratio).
\begin{center}
{\small{\bf Table 1.} Number of kaons with $p_T > 1.8$ GeV/c}\\
\begin{tabular}{lcccccc} \\ \hline\hline
 & & \multicolumn{2}{c}{1-Vertex}  & & 
\multicolumn{2}{c}{2-Vertex} \\
 & &  All  &  No Lepton  & & All  &  No Lepton \\ \hline\hline
Data & & 35.0 & 30.0 & & 
30.0 & 23.0 \\ 
M.C. & & 22.1 & 14.1 & & 
27.5 & 20.3 \\ \hline 
Diff.  & & $12.9\pm 5.9$ & $15.9\pm 5.5$ & & 
$2.5\pm 5.5$ & $2.7\pm 5.5$ 
\\ \hline\hline 
\end{tabular}
\end{center}
At SLD, we select $B$ vertices
that contain an identified $K^\pm$ using the {\v C}erenkov Ring Imaging
Detector. We measure the kaon transverse
momentum w.r.t. the direction defined by the small SLC interaction
point and the well-reconstructed $B$ decay vertex. The signal is enhanced by: 
$i)$ separating the data into a one-vertex 
sample (signal) and a two-vertex sample (control) according to the probability
for all $B$-decay tracks to originate from a single point, 
$ii)$ rejecting decays that contain an identified lepton.
The efficiency for isolating true one-vertex
decays (e.g., charmonium) is estimated at 80\%, whereas only 45\% of standard
$b\rightarrow c$ transitions satisfy the one-vertex requirement.  
We compare the $K^\pm$ transverse momentum 
spectrum observed in the data to that in the Monte Carlo and look for an excess
above 1.8 GeV/c. Our modeling of the $p_T$ resolution is 
cross-checked using leptons whose spectrum is well known.
The results are summarized in Table~1.
We observe in the 1993-95 data ($150k\ Z^0$'s) an excess of 
$12.9\pm 5.9(stat)\pm 3.1(syst)$ decays, without lepton rejection. The 
systematic error is dominated by the uncertainty in the modeling of the $D^0$
momentum spectrum and its two-body decay branching fractions.
The result for the case with lepton rejection is also given in Table 1
(statistical error only). This analysis will be significantly improved with 
the addition of an anticipated large data sample in the near future.
%
% ---- Bibliography ----
%

%
% ---- SLD Author List
%
\newpage
\footnotesize
\centerline{{\large\bf The SLD Collaboration}}
\vskip 0.5cm
\begin{center}
%The SLD author list has been temporarily suppressed in order to save trees.
  \def\iADEL{$^{(1)}$}
  \def\iBOL{$^{(2)}$}
  \def\iBU{$^{(3)}$}
  \def\iBRUN{$^{(4)}$}
  \def\iUCSB{$^{(5)}$}
  \def\iUCSC{$^{(6)}$}
  \def\iCIN{$^{(7)}$}
  \def\iCSU{$^{(8)}$}
  \def\iCOLO{$^{(9)}$}
  \def\iCOL{$^{(10)}$}
  \def\iFER{$^{(11)}$}
  \def\iFRA{$^{(12)}$}
  \def\iILL{$^{(13)}$}
  \def\iLBL{$^{(14)}$}
  \def\iMIT{$^{(15)}$}
  \def\iMASS{$^{(16)}$}
  \def\iMISS{$^{(17)}$}
  \def\iNAG{$^{(18)}$}
  \def\iOREG{$^{(19)}$}
  \def\iPAD{$^{(20)}$}
  \def\iPERU{$^{(21)}$}
  \def\iPISA{$^{(22)}$}
  \def\iRUT{$^{(23)}$}
  \def\iRAL{$^{(24)}$}
  \def\iSOGANG{$^{(25)}$}
  \def\iSOONG{$^{(26)}$}
  \def\iSLAC{$^{(27)}$}
  \def\iTENN{$^{(28)}$}
  \def\iTOH{$^{(29)}$}
  \def\iVAND{$^{(30)}$}
  \def\iWASH{$^{(31)}$}
  \def\iWISC{$^{(32)}$}
  \def\iYALE{$^{(33)}$}
  \def\dead{$^{\dag}$}
  \def\andgen{$^{(a)}$}
  \def\andper{$^{(b)}$}
%
%  \author{                         % author and institution list
%  \baselineskip=.75\baselineskip   % shrink the interline spacing
%
\mbox{K. Abe                 \unskip,\iNAG}
\mbox{K. Abe                 \unskip,\iTOH}
\mbox{T. Abe                 \unskip,\iSLAC}
\mbox{T. Akagi               \unskip,\iSLAC}
\mbox{N.J. Allen             \unskip,\iBRUN}
\mbox{D. Aston               \unskip,\iSLAC}
\mbox{K.G. Baird             \unskip,\iMASS}
\mbox{C. Baltay              \unskip,\iYALE}
\mbox{H.R. Band              \unskip,\iWISC}
\mbox{T. Barklow             \unskip,\iSLAC
% ...
}
\it
  \vskip \baselineskip                   % \bigskip did not work
%  \centerline{(The SLD Collaboration)}   % include collaboration name
  \vskip \baselineskip                   % \bigskip did not work
%
%  }   % end of author list
%
%  \address{                        % institution address list
%  \baselineskip=.75\baselineskip   % shrink the interline spacing
%
  \iADEL
     Adelphi University,
     Garden City, New York 11530 \break
  \iBOL
     INFN Sezione di Bologna,
     I-40126 Bologna, Italy \break
  \iBU
     Boston University,
     Boston, Massachusetts 02215 \break
  \iBRUN
     Brunel University,
     Uxbridge, Middlesex UB8 3PH, United Kingdom \break
  \iUCSB
     University of California at Santa Barbara,
     Santa Barbara, California 93106 \break
  \iUCSC
     University of California at Santa Cruz,
     Santa Cruz, California 95064 \break
  \iCIN
     University of Cincinnati,
     Cincinnati, Ohio 45221 \break
  \iCSU
     Colorado State University,
     Fort Collins, Colorado 80523 \break
  \iCOLO
     University of Colorado,
     Boulder, Colorado 80309 \break
  \iCOL
     Columbia University,
     New York, New York 10027 \break
  \iFER
     INFN Sezione di Ferrara and Universit\`a di Ferrara,
     I-44100 Ferrara, Italy \break
  \iFRA
     INFN  Lab. Nazionali di Frascati,
     I-00044 Frascati, Italy \break
  \iILL
     University of Illinois,
     Urbana, Illinois 61801 \break
  \iLBL
     Lawrence Berkeley Laboratory, University of California,
     Berkeley, California 94720 \break
  \iMIT
     Massachusetts Institute of Technology,
     Cambridge, Massachusetts 02139 \break
  \iMASS
     University of Massachusetts,
     Amherst, Massachusetts 01003 \break
  \iMISS
     University of Mississippi,
     University, Mississippi  38677 \break
  \iNAG
     Nagoya University,
     Chikusa-ku, Nagoya 464 Japan  \break
  \iOREG
     University of Oregon,
     Eugene, Oregon 97403 \break
  \iPAD
     INFN Sezione di Padova and Universit\`a di Padova,
     I-35100 Padova, Italy \break
  \iPERU
     INFN Sezione di Perugia and Universit\`a di Perugia,
     I-06100 Perugia, Italy \break
  \iPISA
     INFN Sezione di Pisa and Universit\`a di Pisa,
     I-56100 Pisa, Italy \break
  \iRUT
     Rutgers University,
     Piscataway, New Jersey 08855 \break
  \iRAL
     Rutherford Appleton Laboratory,
     Chilton, Didcot, Oxon OX11 0QX United Kingdom \break
  \iSOGANG
     Sogang University,
     Seoul, Korea \break
  \iSOONG
     Soongsil University,
     Seoul, Korea 156-743 \break
  \iSLAC
     Stanford Linear Accelerator Center, Stanford University,
     Stanford, California 94309 \break
  \iTENN
     University of Tennessee,
     Knoxville, Tennessee 37996 \break
  \iTOH
     Tohoku University,
     Sendai 980 Japan \break
  \iVAND
     Vanderbilt University,
     Nashville, Tennessee 37235 \break
  \iWASH
     University of Washington,
     Seattle, Washington 98195 \break
  \iWISC
     University of Wisconsin,
     Madison, Wisconsin 53706 \break
  \iYALE
     Yale University,
     New Haven, Connecticut 06511 \break
  \dead
     Deceased \break
  \andgen
     Also at the Universit\`a di Genova \break
  \andper
     Also at the Universit\`a di Perugia \break
\end{center}

\end{document}